# RATING COSMOLOGICAL MODELS

NETA A. BAHCALL

Princeton University Observatory, Princeton, NJ 08544, USA

I summarize the successes and failures of the most popular current cosmological models in accounting for the available observations. This evaluation, presented as the summary of the 11th Potsdam Workshop on Large-Scale Structure, was discussed by all the participants. I present the ratings given to each model based on a confrontation with the data, and the final vote taken.

## 1. Workshop Summary

The workshop summary presented included the following topics:

− Large-Scale Structure
− Clusters of Galaxies
− X-Ray Surveys: The X-Ray Background and Clusters of Galaxies
− Faint Galaxy Counts
− Quasars and Quasar Absorption Lines
− Peculiar Velocities on Large Scales and $\Omega$
− Cosmological Models

Since the proceedings of the meeting contain all the detailed papers presented, I will not repeat the detailed summary here. I only provide the summary of the status of currently popular cosmological models as they confront observations.

## 2. Current Popular Cosmological Models

The current popular cosmological models discussed include the following:

*SCDM*: Standard Cold-Dark-Matter (CDM) [$\Omega = 1$, $h = 0.5$]
Since this model is known to provide a poor match with various observations (see §3), numerous attempts have been made to "patch-up" this standard model by introducing new, ad-hoc free parameters. These non-standard CDM models and other popular models include:

*$\Omega = 1$ Models*
− TCDM: Tilted CDM ["tilt" the spectrum from the standard n=1 slope]
− HCDM: Hot + Cold CDM ["heat" the spectrum by adding neutrinos]
− BSI: Braking Scale Invariance ["break" the initial fluctuation spectrum]

*$\Omega + \Lambda = 1$ Models*
− LCDM: Low-density CDM [add a cosmological constant $\Lambda$]
− $\Lambda$PIB: Primeval Isocurvature Baryonic DM [low-density plus $\Lambda$]

*Ω ~ 0.2 Models*
− OCDM: Open CDM
− OPIB: Open PIB

*Non-Gaussian Models*
− (These models are not discussed here).

Discussions of various aspects of these models are presented in these proceedings by many participants including Bahcall, Borgani, Gottloeber, Hara, Jing, Jones, Kates, Mo, Moscardini, Muller, Pogosyan, Suginohara, Yepes, and more, and references therein. A summary of the comparison with observations is given below.

## 3. Cosmological Models Evaluation

Each model (§2) is contrasted with and evaluated against a set of eleven observational tests. A good or reasonably good fit to the data is scored +1, a bad fit to the data is scored −1, and an intermediate or unclear match is given 0 (or 1/2, as relevant). A question-mark indicates uncertainties. The observational tests listed include most of the relevant tests that are currently available. The tests and the models were discussed throughout the meeting (see proceedings). The observational tests include the following:

− $\Omega_{dyn}$(Mpc): the dynamical $\Omega$ on Mpc scale, determined from the dynamics of galaxy clusters;
− $\Omega_b/\Omega$(cl): the high observed baryon fraction in rich clusters;
− $P_k$(gal): the observed power-spectrum of galaxies;
− MF(cl) the observed mass-function of galaxy clusters;
− $\xi_{cc}$(d): the observed richness-dependent cluster correlation function;
− $\upsilon_p$(1 Mpc): the pairwise peculiar velocity of galaxies at 1 Mpc separation;
− [$\upsilon_p$(50 Mpc)]: the peculiar velocity on large scales, ~ 50$h^{-1}$Mpc (data is inconclusive; see proceedings);
− High z: observations of high redshift objects (damped Ly $\alpha$ lines, quasars, clusters);
− Substructure: substructure in galaxy clusters (no quantified comparison available);
− Inflation: this is *not* a direct observational test; it is included only for the standard "observational" phenomena that inflation can explain;
− Age of Universe: the age of the universe as implied by the age of the oldest stars >13 b.y.).

These observational tests are summarized in Table 1.

Additional critical observational tests include the microwave background fluctuation as a function of angle, $\Delta T/T(\theta)$; the luminosity-function of X-ray clusters; cluster evolution; and the parameter $\beta \equiv \Omega^{0.6}/b$. These tests are not included in the above list since they have either not yet been observationally determined or not yet compared with all the above models. Data on these parameters is currently accumulating and will provide further powerful tests of the models.

Inspection of Table 1 reveals the following problem areas for the different model categories:

$\Omega = 1$ *Models*: these models have problems with the high observed baryon fraction in clusters [$\Omega_b/\Omega(cl)$], the high frequency of high-redshift objects, and with the $> 13$ b.y. age of the universe (especially for $H_o > 50$ km s$^{-1}$ Mpc$^{-1}$). The $\Omega = 1$ models also require a large bias in the distribution of mass versus light.

$\Omega + \Lambda = 1$ *Models*: no significant observational problems. One possible problem may be a high peculiar velocity on large scales, $v_p$(50 Mpc), but the data at present is inconclusive.

$\Omega \sim 0.2$ *Models*: the only potential problems may be the frequent observed substructure in galaxy clusters (but no quantitative comparison is available yet), the loss of inflation (however this is not an observational test; non-standard inflations have also been proposed for open models), and possibly $v_p$(50 Mpc) (but data is inconclusive).

Crucial observations that will help disentangle these model categories include the determination of $\Omega$, $\Lambda$, $H_o$, $P_k$, $\Delta T/T(\theta)$, and the evolution of high-redshift galaxies and clusters. Studies of these tests are all currently underway.

## 4.  Vote on Cosmological Models

All the participants voted on the most likely cosmological model. The question posed was: on which of the above models (Table 1), or an alternate yet unknown model, will you bet your money today? Below is the recorded vote:

| $\Omega = 1$ Models | Vote |
|---|---|
| –HCDM | 4 people |
| –BSI | 4 people |
| $\Omega + \Lambda = 1$ Models | |
| –LCDM | 4 people |
| –$\Lambda$PIB | 0 people |
| $\Omega \sim 0.2$ Models | |
| –Open | 6 people |
| Other Models* | $\gtrsim$ 50 people |

*A newer, more-natural model yet to-be-determined.

The above vote speaks for itself. The prevailing feeling was that most of the current popular models are ad-hoc and that a simpler, more-natural model may be what nature has in store for us.

I wish to thank the Potsdam Workshop organizers, Jan Mucket, Stefan Gottloeber, D.-E. Liebscher and Volker Muller for a stimulating and pleasant meeting.

Table 1: Model Evaluation
(COBE-normalized models)

| OBS. | $\Omega = 1$ | | | | $\Omega + \Lambda = 1$ | | $\Omega \sim 0.2$ | |
|---|---|---|---|---|---|---|---|---|
| | SCDM | TCDM | HCDM | BSI | LCDM | $\Lambda$PIB | OCDM | OPIB |
| $\Omega_{dyn}$(Mpc) | −1 | +1 | +1 | +1[bias] | +1 | +1 | +1 | +1 |
| $\Omega_b/\Omega$(cl) | −1 | −1 | −1 | −1 | +1 | +1 | +1 | +1 |
| $P_k$(gal) | −1 | +1(0) | +1 | +1 | +1 | +1 | +1 | +1 |
| MF(cl) | −1 | +1 | +1 | +1 | +1 | +1 | +1 | +1 |
| $\xi_{cc}(d)$ | −1 | −1 | +1 | +1 | +1 | −1? | +1 | −1? |
| $\upsilon_P$ (1 Mpc) | −1 | +1? | +1? | +1? | +1 | +1 | +1 | +1 |
| ($\upsilon_P$ (50) | +1 | −1 | +1 | +1/2 | −1 | +1 | −1 | +1 |
| High z | +1 | 0 | 0 | 0 | +1 | +1 | +1 | +1 |
| Substructure | +1 | +1 | +1 | +1 | +1? | +1? | 0 | 0 |
| Inflation* | +1 | +1 | +1 | +1 | +1 | +1 | −1 | −1 |
| Age of Univ. | 0 | 0 | 0 | 0 | +1 | +1 | +1 | +1 |
| | [−1 for $\Omega$=1 if $H_o$>50] | | | | | | | |
| Total Score: | −2 | 3 | 7 | 6.5 | 9 | 9 | 6 | 6 |

*Omiting inflation from the list of tests will lower the flat model scores by 1 and increase the open models scores by +1.